# Fourier-transform Ghost Imaging with Hard X-rays

Hong Yu[1], Ronghua Lu[1], Shensheng Han[1,*], Honglan Xie[2], Guohao Du[2], Tiqiao Xiao[2], Daming Zhu[3,4]

[1] *Shanghai Institute of Optics and Fine Mechanics, Chinese Academy of Science, Shanghai, 201800, China*
[2] *Shanghai Institute of Applied Physics, Chinese Academy of Science, Shanghai, 201800, China*
[3] *University of Science and Technology of China, Hefei, 230027, China*
[4] *University of Missouri-Kansas City, Kansas City, Missouri 64110, USA*

**Abstract:** Knowledge gained through X-ray crystallography fostered structural determination of materials and greatly facilitated the development of modern science and technology in the past century. Atomic details of sample structures is achievable by X-ray crystallography, however, it is only applied to crystalline structures. Imaging techniques based on X-ray coherent diffraction or zone plates are capable of resolving the internal structure of non-crystalline materials at nanoscales, but it is still a challenge to achieve atomic resolution. Here we demonstrate a novel lensless Fourier-transform ghost imaging method with pseudo-thermal hard X-rays by measuring the second-order intensity correlation function of the light. We show that high resolution Fourier-transform diffraction pattern of a complex amplitude sample can be achieved at Fresnel region and the amplitude and phase distributions of a sample in spatial domain can be retrieved successfully. The method of lensless X-ray Fourier-transform ghost imaging extends X-ray crystallography to non-crystalline samples, and its spatial resolution is limited only by the wavelength of the X-ray, thus atomic resolution should be routinely obtainable. Since highly coherent X-ray source is not required, comparing to conventional X-ray coherent diffraction imaging, the method can be implemented with laboratory X-ray sources, and it also provides a potential solution for lensless diffraction imaging with fermions, such as neutron and electron where the intensive coherent source usually is not available.

Since Max Laue discovered X-ray diffraction in crystals in 1914, X-ray crystallography has become a powerful tool in exploring and analyzing the internal structures of complex materials, such as biomolecular structures and nanomaterials[1-3]. The resolution of X-ray crystallography is only limited by the wavelength of the X-ray, which provides the opportunity to visualize the atomic details of crystalline structures.

However, structure information of many important molecular materials, such as membrane proteins, is still out of reach, because these materials are difficult to grow into macroscopic crystals. Furthermore, with the rapid development of nanoscience and biology, it has become an urgent need to obtain atomic resolution images of the internal structure of samples in their natural states instead of in crystals. In 1999, Coherent diffraction imaging (CDI) method[4-6] was proposed to extend X-ray crystallography to allow imaging non-crystalline structures in nanoscale by illuminating the samples with coherent X-rays and recording the diffraction patterns in far-field. Fresnel CDI[7,8] and ptychography technique[9,10] were also proposed to circumvent the intrinsic restriction of sample size in classical CDI. Nevertheless, due to the requirements for high coherence and brightness, synchrotron radiation or X-ray free electron laser sources is still essential to X-ray CDI applications, and high resolution imaging using laboratory X-ray source with CDI techniques remains to be achieved.

Most of conventional imaging methods are based on detection of intensity distribution of light fields, i.e. the first-order correlation of the light. In fact, imaging in both real and reciprocal space can be realized with thermal light through ghost imaging technique, a phenomenon first observed in quantum regime two decades ago[11,12], by measuring higher order correlation of light fields. Different from conventional methods, in a typical ghost imaging system, the light field passing through or reflected by a sample is recorded only with a non-spatially resolving detector(i.e. a bucket or point detector), and the sample's information is acquired from the second-order intensity fluctuation correlation of the scattered and unscattered lights. Ghost imaging has been proved and demonstrated with classical visible thermal light[13-19] and applied quickly in remote sensing, photolithography, super resolution imaging, single-pixel three dimensional camera, etc[20-31]. A lensless X-ray Fourier-transform ghost imaging scheme[15,32] with spatially incoherent illumination, where the Fourier-transform diffraction pattern of the sample can be acquired at Fresnel region, had also been proposed to achieve the same diffraction pattern as in CDI. Therefore Fourier-transform ghost imaging provides a possibility to achieve atomic resolution images of non-crystalline samples with widely accessible laboratory incoherent X-ray sources. However, a beam splitter is needed in the scheme to generate two copies of the incident light field, which is difficult for X-ray.

---

[*] sshan@mail.shcnc.ac.cn

In this paper, we report for the first time the success of an experiment which demonstrates the feasibility of performing hard X-ray Fourier-transform ghost imaging (FGI) using a navel experimental approach. The experiment was based on the same principle of visible light FGI which is illustrated in Fig.1(a). A light beam from a spatially incoherent source is divided into two beams, a testing beam and a reference beam, after passing through a beam splitter. The light in the testing beam passes through a sample at a distance $d_1$ and then is recorded by a point detector $D_t$ positioned at a distance $d_2$ from the sample. The reference beam does not pass through the sample at all, but its intensity distribution after passing through a distance d is recorded by a spatially resolved panel detector $D_r$.

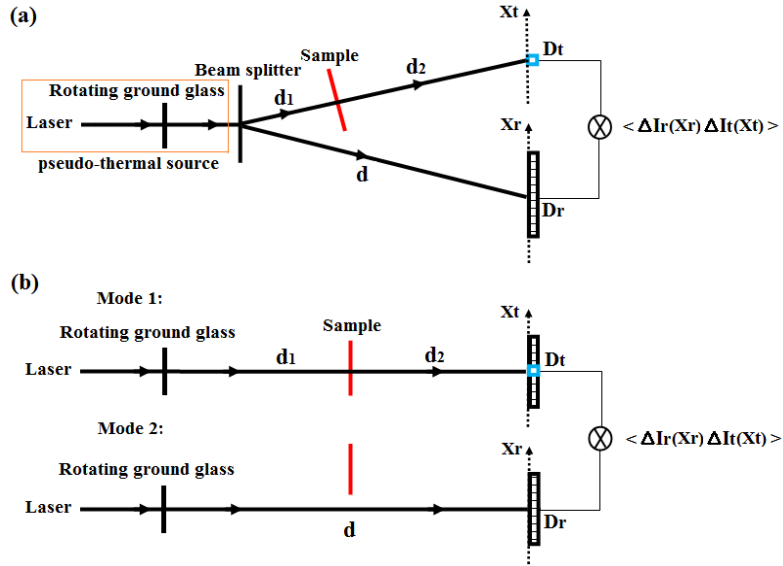

Fig. 1. Illustration of the experimental setup for FGI without a beam splitter. (a) is the principle of lensless FGI, and (b) is the setup of a lensless FGI system without a beam splitter: For a controllable pseudo-thermal light source, when the sample is inserted into the light beam as in mode 1, each pixel of the panel detector can be considered as a point detector $D_t$ in the testing beam of FGI, while when the sample is moved out of the beam as in mode 2, the panel detector records the intensity distribution without passing through the sample as the reference beam detector $D_r$ of FGI, if only the two mode measurements are completed in one pseudo-coherent duration determined by the rotating speed of the ground glass.

As the derivations detailed in reference [15] show that, under the condition where $d_1 + d_2 = d$, the correlation function between the intensity fluctuation of reference beam $\Delta I_r(x_r)$ and that of the testing beam $\Delta I_t(x_t)$ measured by the two detectors is directly related to the modulus of the Fourier transformation of the sample's transmittance, and such a relation can be expressed as

$$\langle \Delta I_r(x_r) \Delta I_t(x_t) \rangle = \frac{I_0^2}{\lambda^4 d_2^4} \left| T\left(\frac{2\pi(x_t - x_r)}{\lambda d_2}\right) \right|^2, \quad (1)$$

where $\Delta I_k(x_k) = I_k(x_k) - \langle I_k(x_k) \rangle$, $k = r$ or $t$, $x_r$ and $x_t$ are the coordinates at the detector planes in the reference and testing paths respectively, $\lambda$ is the wavelength, $I_0$ is the intensity of the incident light, $T\left(\frac{2\pi(x_t - x_r)}{\lambda d_2}\right)$ is the Fourier transformation of the sample's transmittance. Real space image of the sample can be retrieved from $\left| T\left(\frac{2\pi(x_t - x_r)}{\lambda d_2}\right) \right|$ as in conventional X-ray CDI. Thus, the sample can be imaged by recording the incoherent spatial intensity distribution of the beam without passing through the sample and then correlated with the signal detected by a single pixel detector placed behind the sample. Such an imaging method removes the need for a beam stop and avoids missing low-frequency data in the diffraction patterns as in traditional X-ray CDI, which has been intensively studied using the visible pseudo-thermal light source in the past decade[17,33].

However, to test the FGI in the hard x-ray regime is challenging. For hard X-rays, there is no perfect beam splitter which can produce twin beams as the case of visible light[32]. To circumvent such a difficulty, we notice that the condition $d_1 + d_2 = d$ is satisfied in the scheme. So, for controllable pseudo-thermal light source, rather than splitting the thermal light into two beams as in Fig.1(a), we can use an equivalent scheme as shown in Fig.1(b), which uses only one spatially incoherent pseudo-thermal light beam and a fixed panel detector by shuttling the sample in and out of the beam in one pseudo-coherent duration (during which the light source is stable). When the sample is inserted into the beam, the signals detected by a single pixel of the panel detector serve as that detected in the point detector in the testing beam, when the sample is moved out of the beam, the signals detected serve as that detected in the panel detector in the reference beam, so both the intensity fluctuation $\Delta I_t(x_t)$ and $\Delta I_r(x_r)$ can be acquired by the same panel detector if only the two measurements are completed in one pseudo-coherent duration.

The experiment was performed on the 13W beamline at Shanghai Synchrotron Radiation Facility (SSRF) which is dedicated to X-ray imaging and biomedical applications. Fig.2(a) shows the experimental setup. A pseudo-thermal X-ray source[34-36], which can generate a controllable chaotic X-ray speckle pattern fluctuation to emulate the behavior of a spatially incoherent X-ray source, was used to illuminate the sample. The pseudo-thermal X-ray source is produced by a monochromatic X-ray beam passing through a slit array and a moveable gold film deposited on a Si3N4 substrate. The monochromatic X-ray beam was produced by passing the X-rays emitted from the synchrotron source through a double crystal monochromator with an energy resolution $\Delta E/E \approx 10^{-3}$. The flux of the x-ray was $3 \times 10^{10}$ photons/mm²/sec and the energy was centered at 12.1 keV(0.1 nm wavelength). The slit array is positioned in the optical path of the X-ray beam, and each slit of the array has a dimension less than or equal to the X-ray's coherent area at the slit array plane (50um×10um), so that each of the transmitted X-ray sub-beam passing through a slit is spatially coherent. The gold film with randomly distributed holes of diameter<1um was mounted on a computer-controllable translational device and placed closely behind the slit array, the depth of the holes are designed to be $\lambda/2(n-1) = 2.7um$ to form a phase difference of $\pi$ between the area with and without holes for 12.1 keV hard X-ray. After the bundle of spatially coherent X-ray sub-beams from the slit array passing through the gold film, chaotic distributed X-ray speckle patterns are produced because of the spatially stochastic interference of the randomly modulated spatial coherent X-ray sub-beams from the gold film. The size of the gold film is much larger than the whole beam cross-section of the monochromatic X-rays, thus when the gold film is moved transversely by the translational device to make the different part of the film be illuminated, a pseudo-thermal X-ray beam with chaotic fluctuating intensity distributions was produced and served as the incoherent X-ray source in our experiment. The CCD camera with effective pixel size of 0.37um*0.37um was placed 43cm downstream from the gold film. The experimental sample was placed on a controllable translational stage, which can move the sample in and out of the pseudo-thermal X-ray beam as shown by the red arrow in Fig.2(a), for performing $\Delta I_t(x_t)$ and $\Delta I_r(x_r)$ measurements. In one measurement, a pair of X-ray signals, i.e. the intensity fluctuation of the testing beam $\Delta I_t(x_t)$ and that of the reference beam $\Delta I_r(x_r)$, was acquired as shown in Fig.2(b), and the gold film was trigged to move transversely again after each of the measurements was completed. We should mention that the drift of the sample position relative to the illuminating pseudo-thermal X-ray beam introduced by the shuttling was almost inevitable, but it makes little influence to FGI system as demonstrated theoretically and experimentally in reference [37].

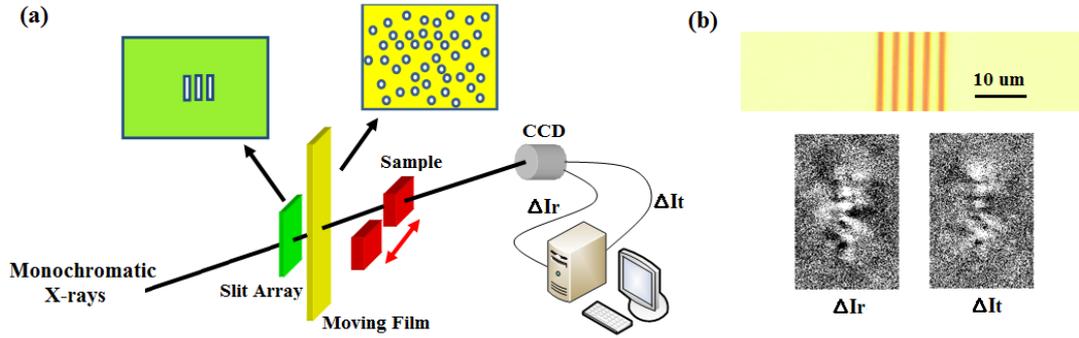

Fig. 2. Experimental setup for X-ray FGI using a pseudo-thermal X-ray source is shown in (a). A monochromatic X-ray beam passes through a slit array and a moving film to generate a controllable pseudo-thermal X-ray speckle pattern fluctuation which is used in the experiment as the pseudo-thermal X-ray source. The sample is moved in and out of the beam to get the intensity fluctuation $\Delta I_t(x_t)$ and $\Delta I_r(x_r)$ respectively as required by the scheme of FGI without a beam splitter. Fig. 2(b) shows the optical microscope image of the sample and an example of intensity distribution pattern pairs $\Delta I_t(x_t)$ and $\Delta I_r(x_r)$ acquired in X-ray FGI experiment.

The sample in our experiment was a 2.2um thick gold film with five slits on a Si3N4 substrate. The slits were separated by $d_{slit} = 3um$ and the width of each slit was 1um. Fig.2(b) shows the optical microscope image of the sample. Since the wavelength of the pseudo-thermal X-ray source was 0.1nm, the intensity attenuation and phase difference between the slits and the surrounding gold area was 53% and $0.82\pi$, respectively. The distance from sample to CCD camera is 33cm and the width of the sample illuminated by the 0.1nm pseudo-thermal X-rays is 13um, so the far-field diffraction condition ($D^2/\lambda = 13um^2/0.1nm = 1.69m$, $D$ is the total width of the five slits) is not satisfied. For FGI purpose, signals from only one fixed single pixel in the intensity pattern, recorded while the imaging example was placed in the beam, were needed to construct the correlation function. For convenience, we recorded the entire intensity pattern of the beam passing through the sample. An example of the intensity distribution pattern pairs is shown in Fig. 2(b). Obviously, the X-ray intensity patterns display featureless random distributions, and no diffraction patterns of the sample can be directly observed.

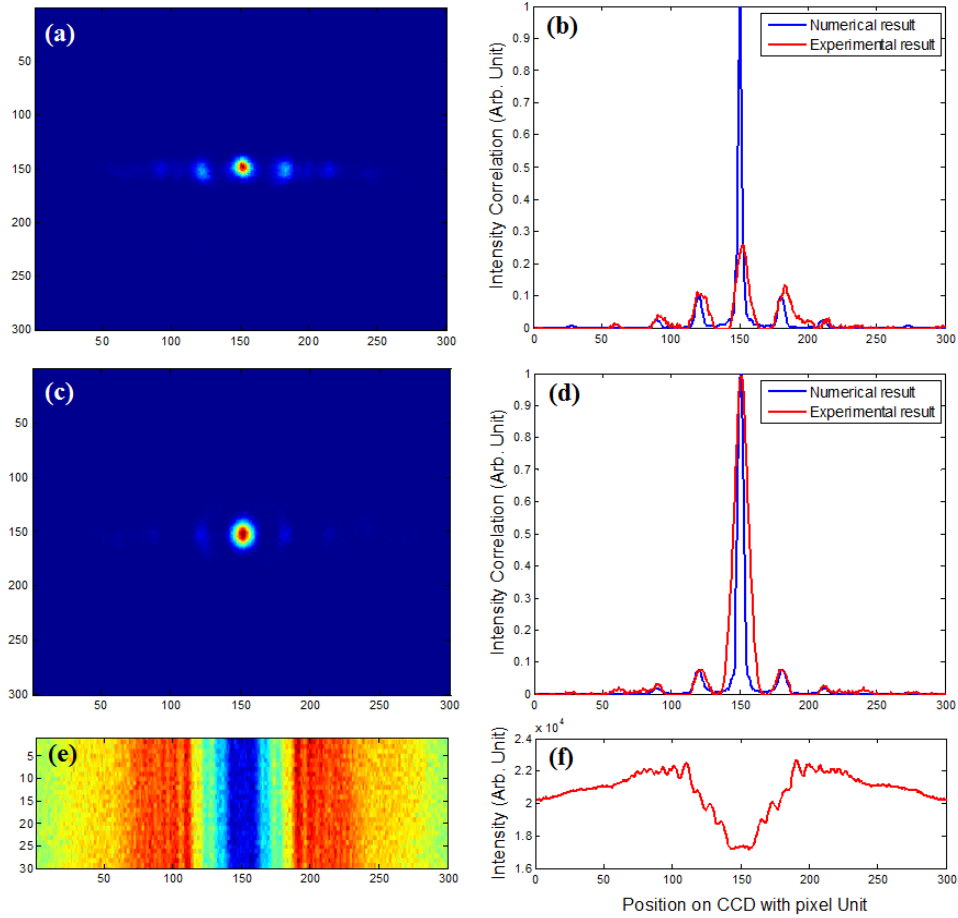

Fig. 3. Diffraction patterns of the sample obtained with X-rays of 0.1nm wavelength. (a) is the Fourier-transform diffraction pattern of the sample's transmittance obtained by X-ray FGI, (c) is the Fourier-transform diffraction pattern of the squared modulus of the sample's transmittance obtained in X-ray FGI, (e) is the intensity distribution obtained by illuminating the sample directly with synchrotron X-rays. The red lines in (b), (d) and (f) are the cross-section curves of (a), (c) and (e), respectively. The blue lines in (b) and (d) are the corresponding numerical results obtained by Fraunhofer diffraction integral.

The Fourier-transform diffraction pattern of the complex amplitude sample was obtained by calculating the correlation between the intensity fluctuations of the testing beam $\Delta I_t(x_t)$ and the reference beam $\Delta I_r(x_r)$ following Eq.(1). However, to improve the sampling efficiency in the calculation, we applied a reconstruction algorithm making use of sparsity constraints of image[38] in reconstructing the Fourier-transform diffraction patterns. Fig.3(a) is the sample's diffraction pattern obtained by X-ray FGI with 284 pairs of measurements data used in reconstruction calculation. The cross-section curve of Fig.3(a) is shown by the red line in Fig.3(b), and the peak spacing of the red line in Fig.3(b) is about 11.1um(0.37um/pixel*30 pixels=11.1um), which is in agreement with the theoretical value of the peak spacing($\lambda d_2 / d_{slit} = 0.1nm \times 33cm / 3um = 11um$) predicted by Eq.(1). The blue line in Fig.3(b) shows the numerical result of the sample's Fourier transformation, and it agrees well with the experiment result. Therefore, the sample's Fourier-transform diffraction pattern was obtained at Fresnel region by X-ray FGI, which is different from the case in X-ray CDI, where the sample's Fourier-transform diffraction pattern should be obtained at far-field.

By calculating the auto-correlation of the intensity fluctuation of the testing beam $\Delta I_t(x_t)$, the Fourier-transform diffraction pattern of the squared modulus of the sample's transmittance can also be obtained[39]. The Fourier-transform diffraction pattern of the squared modulus of the sample's transmittance obtained in X-ray FGI and the corresponding cross-section curves are shown in Fig.3(c) and Fig.3(d), respectively. As Fig.3(d) shows, the experiment result agrees well with the numerical Fourier transformation result. Thus, the amplitude and phase information of the complex amplitude sample were obtained separately in our X-ray FGI experiment.

For reference, the X-ray intensity distribution recorded by the CCD camera when the sample was directly illuminated by the monochromatic X-ray beam emitted from the synchrotron source through the double crystal monochromator is shown in false-color representation as Fig.3(e). Fig.3(f) shows the cross-section curve of Fig.3(e). By comparing Fig.3(e) with Fig.3(a) and Fig.3(c), it can be found that the pattern obtained when illuminating the sample with monochromatic X-rays is apparently different from the Fourier-transform patterns obtained in X-ray FGI.

Using a two-step phase-retrieval image reconstruction process based on FGI[42], in which firstly the amplitude part of the sample's transmittance is retrieved from the Fourier-transform patterns in Fig.3(c), then combining the retrieved amplitude part of the sample's transmittance with the Fourier-transform pattern in Fig.3(a), the phase part of the sample's transmittance was retrieved. The retrieved amplitude and phase distributions of the sample's transmittance are shown in Fig.4(a) and Fig.4(b), respectively. The maximum spatial frequency used in the reconstruction can be calculated as $q_{\max} = \dfrac{0.37 um/pixel * 300\, pixels}{2\lambda d_2}$, so the pixel size in the retrieved image is $\dfrac{1}{2q_{\max}} = 0.297 um$, and the separate distance between the slits in the retrieved image is 2.97um(0.297um/pixel*10 pixels), which is in completely agreement with the spatial feature of the sample.

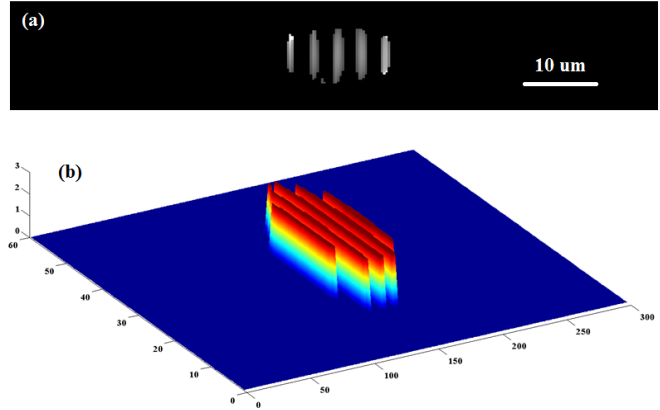

Fig. 4. The sample's amplitude distribution (a) and phase distribution (b) in spatial domain retrieved from the Fourier-transform diffraction patterns obtained in X-ray FGI.

Our experimental results demonstrated for the first time that Fourier-transform ghost imaging can be achieved using pseudo-thermal hard X-rays. The diffraction patterns are qualified enough to retrieve the amplitude and phase distributions of the sample in spatial domain. The spatial resolution of FGI with incoherent X-rays is determined by the maximum spatial frequency of the Fourier-transform diffraction pattern, which means the spatial resolution of lensless X-ray FGI is only limited by the wavelength, and provides the potential to achieve atomic resolution images of non-crystalline samples with laboratory X-ray sources.

In summary, we have experimentally demonstrated Fourier-transform ghost imaging with pseudo-thermal hard X-rays and high resolution Fourier-transform diffraction pattern of the sample has been achieved at Fresnel region by measuring the second-order intensity correlation of the lights, the amplitude and phase distributions of the sample in spatial domain have been retrieved successfully. This method extends X-ray crystallography to non-crystalline samples and, as been a lensless imaging scheme, the spatial resolution of X-ray FGI is only limited by the wavelength. An important feature of X-ray FGI method is that it does not rely on highly coherent X-ray source to realize X-ray diffraction imaging, therefore it provides a feasible way to achieve atomic resolution images of non-crystalline samples with widely accessible laboratory X-ray sources. Eliminating the need for intensive coherent source, FGI of not only bosons but also fermions such as neutron and electron, can also be expected, so it provides a glimpse of possibility of revolutionizing the current neutron and electron scattering methods widely used in researches in materials sciences as well as biomedicine. Furthermore, the Fourier-transform diffraction pattern is acquired from correlated calculations, which removes the need for a beam stop and avoids missing low-frequency data in the diffraction patterns as in traditional X-ray CDI. Finally, the high frequency portion of the diffraction pattern that is out of the CCD detection area can also be captured by using the intensity distribution patterns deviate from the center point in pairs[39], which means the spatial resolution of X-ray FGI systems can be doubly enhanced.

The work is supported by the Hi-Tech Research and Development Program of China under Grant Projects No.2013AA122901 and No.2013AA122902, the National Natural Science Foundation of China under Grant Projects No.11105205, and the Shanghai Fundamental Research Project No.09JC1415000.